\documentclass{article}

\usepackage{arxiv}

\usepackage[utf8]{inputenc} % allow utf-8 input
\usepackage[T1]{fontenc}    % use 8-bit T1 fonts
\usepackage{hyperref}       % hyperlinks
\usepackage{url}            % simple URL typesetting
\usepackage{booktabs}       % professional-quality tables
\usepackage{amsfonts}       % blackboard math symbols
\usepackage{nicefrac}       % compact symbols for 1/2, etc.
\usepackage{microtype}      % microtypography
\usepackage{lipsum}
\usepackage{graphicx}
\graphicspath{ {./images/} }

\usepackage{amsmath}
\usepackage{amssymb}
\usepackage{multirow}
\usepackage{cleveref}
\usepackage{enumitem}
\usepackage{makecell}
\usepackage{color, colortbl, xcolor}
\usepackage{listings}
\usepackage{fancybox}
\usepackage{float}
\definecolor{Gray}{gray}{0.92}

\title{PixRec: Leveraging Visual Context for Next-Item Prediction in Sequential Recommendation}

\author{
 Sayak Chakrabarty $^\dagger$\\
  Department of Computer Science,
  Northwestern University\\
  Evanston IL 60208, USA \\
  \texttt{sayakchakrabarty2025@u.northwestern.edu} \\
   \And
 Souradip Pal $^\dagger$\\
  Elmore Family School of Electrical and Computer Engineering\\
  Purdue University\\
  West Lafayette IN 47906, USA \\
  \texttt{pal43@purdue.edu} \\
}

\begin{document}
\maketitle

% Manually create the footnote text
% The \def\thefootnote command temporarily redefines how footnote markers appear
\def\thefootnote{\dag}\footnotetext{These authors contributed equally to this work.}
% Reset the footnote counter for subsequent notes (if any)
\def\thefootnote{\arabic{footnote}}

\begin{abstract}
Large Language Models (LLMs) have recently shown strong potential for usage in sequential recommendation tasks through text-only models, which combine advanced prompt design, contrastive alignment, and fine-tuning on downstream domain-specific data. While effective, these approaches overlook the rich visual information present in many real-world recommendation scenarios, particularly in e-commerce. This paper proposes \textit{PixRec} - a vision-language framework that incorporates both textual attributes and product images into the recommendation pipeline. Our architecture leverages a vision–language model backbone capable of jointly processing image–text sequences, maintaining a dual-tower structure and mixed training objective while aligning multi-modal feature projections for both item--item and user--item interactions. Using the Amazon Reviews dataset augmented with product images, our experiments demonstrate $3\times$ and $40$\% improvements in top-rank and top-10 rank accuracy over text-only recommenders respectively, indicating that visual features can help distinguish items with similar textual descriptions. Our work outlines future directions for scaling multi-modal recommenders training, enhancing visual–text feature fusion, and evaluating inference-time performance. This work takes a step toward building software systems utilizing visual information in sequential recommendation for real-world applications like e-commerce.

\end{abstract}

% keywords can be removed

\section{Introduction}
Recommender systems aim to learn from past user--item interactions in order to surface the most relevant next item. Beyond traditional collaborative filtering and static embeddings, sequential recommenders explicitly model temporal structure in user behavior to predict the next interaction from a chronologically ordered history. Recently, large language models (LLMs) have emerged as promising backbones for sequential recommendation because they naturally model long-range dependencies and can operate directly over rich textual attributes. SASRec~\cite{8594844}, BERT4REC~\cite{10.1145/3357384.3357895}, Recformer~\cite{10.1145/3580305.3599519}, GPT4Rec~\cite{li2023gpt4rec} are strong representatives of this trend: CALRec~\cite{li2024calrec} frames next-item prediction as text generation with carefully designed prompts, augments the language modeling loss with contrastive alignment at the user/item level, and uses a two-stage fine-tuning paradigm (multi-category joint training followed by category-specific adaptation) together with a practical BM25 retrieval layer to map generated text back to concrete catalog items.

However, many real-world domains, particularly e-commerce, are fundamentally \emph{multi-modal}. Two items can share near-identical textual attributes  like title, category, brand, etc., while differing in visual attributes like style, colorway, packaging etc. that strongly influence preference. Relying on text alone can thus hamper disambiguation, especially at the top of the ranking where fine-grained feature cues matter most. Conversely, images alone are insufficient for robust recommendations without semantic context and the sequential patterns encoded in language. This motivates extending text-only LLM recommenders to vision-language models (VLMs) that jointly reason over images and text within the same generative framework.

This paper proposes a multi-modal extension of text-based recommenders that integrates product images alongside flattened textual attributes for sequential recommendation. Concretely, we adapt a vision--language causal modeling approach to ingest \emph{text+image} sequences for the user history and target item, while employing the following techniques: (i) a prompt with a user-history prefix and repeated per-item prefix to cue next-item generation; (ii) a mixed training objective that combines next-item generation with contrastive alignments; and (iii) a retrieval layer that ranks catalog items using quasi round-robin BM25 \cite{robertson2009probabilistic} over the generated candidates. Unlike earlier approaches, our focus is on the \emph{multi-modal} setting with parameter-efficient fine-tuning (PEFT) for practical training on modest hardware.

Our contributions can be summarized as follows:
\begin{itemize}
    \item \textbf{Image-text Processing Data Pipeline:} Our methods provide a pipeline to construct sequential datasets from recent Amazon Review categories with paired images and flattened text attributes, and generate a catalog corpus compatible with BM25 ranking.
    \item \textbf{Practical PEFT Recipe for Fine-tuning:} Our technique provides a parameter-efficient training setup (QLoRA) that enables stable fine-tuning of recommendation models on a single commodity GPU.
    \item \textbf{Multi-modal Recommendations:} Our framework to jointly process images and text for sequential recommendation tasks, retaining the user/target two-tower structure with contrastive-alignment objective while adapting input formatting and projection heads to extract vision and text features, thus illustrating the potential effectiveness of visual information on next-item prediction.
\end{itemize}
In what follows, we detail our architecture, datasets, fine-tuning strategy, retrieval procedures, and initial results, and subsequently outline a road-map for scaling multi-modal training, improving fusion, and evaluating recommendation performance for real-world e-commerce platforms.

\section{Related Work}
Traditional sequential recommendation systems have evolved from ID-only models such as SASRec~\cite{8594844} and Bert4Rec~\cite{10.1145/3357384.3357895} to text-augmented approaches like FDSA~\cite{ijcai2019p600}, S3-Rec~\cite{10.1145/3340531.3411954} and UniSRec~\cite{10.1145/3534678.3539381}. Another thread of research explored deriving IDs from attribute embedding~\cite{rajput2023recommender,10.1145/3640457.3688190,chakrabarty2025time,banerjee2025clt,10.1145/3543507.3583434}, which mitigates some of the challenges associated with random IDs, albeit at the cost of increased system complexity.

With the advent of pretrained LLMs, which include encoder-only models like BERT~\cite{devlin-etal-2019-bert,10.1145/3580305.3599519}, encoder-decoder models like T5~\cite{10.5555/3455716.3455856}, and decoder-only models like PaLM~\cite{10.5555/3648699.3648939,anil2023palm}, GPT~\cite{brown2020language}, and LLama~\cite{2023arXiv230213971T}, researchers explored leveraging these backbones in sequential recommendation tasks, harnessing the extensive `knowledge' encoded within their parameters. GPT4Rec~\cite{li2023gpt4rec} fine-tunes a pre-trained GPT-2 model to generate the next-item titles. CALRec~\cite{li2024calrec} builds on these advances by leveraging autoregressive LLMs in a two-tower architecture, combining next-item generation loss with two contrastive losses---target-on-target ($\mathcal{L}_{TT}$) and user-on-target ($\mathcal{L}_{UT}$)---to align user and item representations in a shared space. A quasi-round-robin BM25 retrieval method matches generated candidates to the item corpus.
 
Beyond textual modeling as in \cite{NIPS2017_3f5ee243,svete2024transformers,chakrabarty2025readmeready}, multi-modal systems incorporate additional modalities such as images~\cite{zhai2024actions,chakrabarty2024mm,chen2019personalized}. Images supply discriminative signals \cite{DBLP:conf/iclr/DosovitskiyB0WZ21} that are often absent or under-specified in titles and categories (e.g., design variants, covers, bundles). In a generative pipeline, these visual cues can steer the models toward more specific completions and yield higher top-rank precision after retrieval. Moreover, aligning user-history representations with target-item representations at the visual--text feature level provides an additional path for contrastive learning to regularize generation and improve candidate quality. Early multi-modal methods fused visual and textual embeddings using CNN or attention-based architectures to capture complementary signals. More recently, vision--language models (VLMs) such as CLIP~\cite{radford2021learningtransferablevisualmodels} and PaliGemma2~\cite{paligemma2} have shown strong performance in cross-modal understanding and retrieval, making them attractive backbones for multi-modal recommendation. 

Integrating multi-modal inputs into LLMs for recommendation is relatively unexplored. Some recent works adapt LLMs with image encoders for product search, visual question answering, document understanding, automatic code generation from UI designs \cite{10.1145/3701716.3717567,lu2022learn,bolonkin2024judicial,wang-etal-2024-docllm,si2024design2code,zhang2023sockdef} but there is limited work on multi-modal \emph{sequential} recommendation with auto-regressive generation and contrastive alignment. Our work fills this gap by extending purely textual paradigm to handle image--text sequences.

\section{Methods and Experimental Results}
Here, we have a set of users $U$, where a user $u \in U$ is associated with an interaction sequence comprising of $n$ items $(t_1, t_2, ..., t_n)$ arranged in chronological order (note that the value of $n$ may vary across different users). The sequential recommendation task is then defined as follows: given the user’s historical interaction sequence $t_{1:n-1} = (t_1, t_2, ..., t_{n-1})$, a sequential recommender aims to predict the target item $t_n$ , the last item in the user's full interaction sequence. Our ultimate aim is to exploit the rich visual information along with textual context inherent in real-world data and to capitalize on VLMs' strong language understanding and reasoning capabilities for next item prediction.

\subsection{Model Architecture}
We adapt a two-tower design in our vision--language setting.
\begin{itemize}
    \item \textbf{History+Target Tower:} The concatenated user history and the target item (each as text+image placeholders with aligned pixel tensors) are fed to the backbone to obtain hidden states. Mean pooling over token positions before and after a learned separator yields the \emph{user history} representation $\mathbf{v}^U$ and the \emph{user history-conditioned target item} representation $\mathbf{v}^{T|U}$, respectively.
    \item \textbf{Target-Only Tower:} The target item (text+image) alone is encoded to obtain the target item representation $\mathbf{v}^T$.
\end{itemize}
Two linear heads project the hidden states generated from the backbone model into a lower-dimensional space for contrastive alignment. The language modeling (LM) head of the backbone provides next-token logits for next-item generation.

\begin{figure}[ht!]
    \centering
    \includegraphics[width=0.8\linewidth]{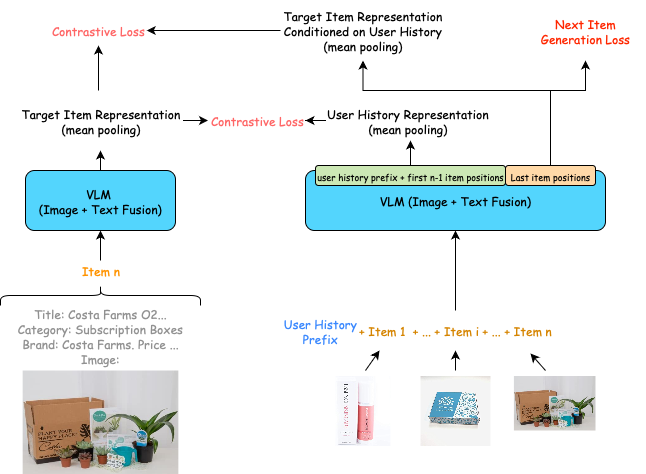}
            \caption{PixRec: Training Framework.}
            \label{fig:framework}
\end{figure}

\paragraph{Backbone Variants}
We experimented with two VLM backbones:
\begin{itemize}
    \item \textbf{SmolVLM} (\texttt{SmolVLM-256M-Instruct}): A compact image--text causal VLM from HuggingFace Smol Models Research, showing strong performance on multimodal tasks, suitable for fast iteration and low-memory training.
    \item \textbf{PaliGemma2} (\texttt{paligemma2-3b-mix-224}): A stronger image--text causal VLM from Google with improved visual grounding, used as our primary model for fine-tuing using PEFT.
\end{itemize}

\begin{table}[!htbp]
\centering
\caption{Hyper-parameters \& PEFT configuration used in the baseline training and fine-tuning of vision-language models on Amazon Reviews dataset}
\label{table:hyperparams}
\begin{tabular}{cc}
    \toprule
    \textbf{Hyperparameter} & \textbf{Value} \\
    \midrule
    \texttt{projection\_dim} & 128 \\
    \texttt{rank} & 8 \\
    \texttt{lora\_alpha} & 8 \\
    \texttt{lora\_dropout} & 0.1 \\
    \texttt{learning\_rate} & 0.0001 \\
    \bottomrule
\end{tabular}
\end{table}

\subsection{Contrastive Objective}
We used a mixed objective for aligning text+image features from both historical user trend and items. The objective consists of the following components:
\begin{itemize}
    \item \textbf{Next-Item Generation (NIG)}: The item-level cross-entropy loss computed only on the target item given the items in the user's interaction history.

\begin{equation}
\label{formula:NIP}
\mathcal{L}_{NIG}=\mathop{\mathbb{E}}\limits_{\textsc{t} \sim p(\textsc{t})} \left[\sum_{j=m+1}^{l}\text{log}\, P(t_{j}|t_{1:j-1};\theta) \right]
\end{equation}

where $p(T)$ denotes the data distribution such that the tokenized user sequence (for a user) is given by $T = (t_1, t_2, ..., t_l)$, where $l$ is the tokenized sequence length, and $\theta$ denotes the trainable parameters.
    \item \textbf{Contrastive Alignments:} InfoNCE \cite{oord2018representation} losses between $(\mathbf{v}^{T|U}, \mathbf{v}^{T})$ \& $(\mathbf{v}^{U},\mathbf{v}^{T})$ with contrastive temperature $\tau_c{=}0.5$.
    \begin{align}
\begin{split}\label{formula:CL_TT}
\mathcal{L}_{TT}=&-\frac{1}{N_{b}}\!\sum_{i=1}^{N_{b}}\text{log} \frac{\text{exp}(\text{cos}(\mathbf{v}^{T|U}_{i},\mathbf{v}^{T}_{i})/\tau_{c})}{\sum_{j=1}^{N_{b}}\text{exp}(\text{cos}(\mathbf{v}^{T|U}_{j},\mathbf{v}^{T}_{i})/\tau_{c})},
\end{split}\\
\begin{split}\label{formula:CL_UT}
\mathcal{L}_{UT}=&-\frac{1}{N_{b}}\!\sum_{i=1}^{N_{b}}\text{log} \frac{\text{exp}(\text{cos}(\mathbf{v}^{U}_{i},\mathbf{v}^{T}_{i})/\tau_{c})}{\sum_{j=1}^{N_{b}}\text{exp}(\text{cos}(\mathbf{v}^{U}_{j},\mathbf{v}^{T}_{i})/\tau_{c})},
\end{split}
\end{align}
\end{itemize}
The final objective becomes
\begin{equation}
\mathcal{L} \;=\; (1-\alpha-\beta)\,\mathcal{L}_{\text{NIG}} \;+\; \alpha\,\mathcal{L}_{TT} \;+\; \beta\,\mathcal{L}_{UT},
\end{equation}
with $\alpha{=}0.125$ and $\beta{=}{-}0.025$.

\subsection{Supervised Fine-Tuning}
\label{sec:peft}
Since multi-modal models are large, fully training all parameters is quite compute and memory intensive and may overfit when data are limited. \emph{Parameter-Efficient Fine-Tuning} (PEFT) addresses this by training a small number of additional parameters while freezing most of the base model. Hence, we use LoRA-style adapters \cite{hu2022lora} with the configuration as shown in Table \ref{table:hyperparams}. We wrap our backbone with \texttt{PeftModel} and train only the adapter weights plus our user and item projection heads. The projection heads are trained fully in the traditional way without using the adaption technique. Our implementation include using an \texttt{SFTTrainer} wrapper that computes the contrastive mixed loss per batch and optimizes the objective using an \texttt{AdamW} optimizer.

\subsection{Data Pipeline and Preprocessing}
We use the Amazon Reviews 2023 dataset \cite{hou2024bridging} for our fine-tuning process. The details of each of the review category used are shown in \ref{table:datasets}. Multi-category fine-tuning was performed for each model where both \textit{Subscription\_Boxes} and \textit{Magazine\_Subscriptions} categories were mixed to prepare a multi-category dataset consisting of $7973$ example sequences and then evaluated only on \textit{Subscription\_Boxes} category. From each item's metadata, we extract \texttt{Title}, \texttt{Category}, \texttt{Brand}, \texttt{Price} and the ``large'' \texttt{Image} (from the \emph{MAIN} product variant). We discard items without an image to ensure every sample has image--text pairs. Next, the reviews were grouped by \texttt{user\_id}, sort by \texttt{timestamp}, to build per-user sequences. Given the sparsity of image coverage, we set the minimum length to $\ge 2$ interactions per user and randomly sample 2 to 6 interaction for each case to ensure enough multi-modal examples. Each item features were subsequently flattened into texts as follows:
\begin{quote}
\small
\texttt{Title: <title>. Category: <category>. Brand: <brand>. Price: <price>. Image: <image>}
\end{quote}
Using these features, we build an item corpus which was later used for BM25 matching during evaluation for accurately selecting the top relevant items from the generated text.

\begin{table}[ht]
\small
\centering
\caption{Amazon Reviews Dataset categories showing the number of Users, Items and Ratings}
\label{table:datasets}
\begin{tabular}{lcccc}
\toprule
\textbf{Category}  & \textbf{Users} & \textbf{Items} & \textbf{Ratings} \\
\midrule
\textit{Subscription\_Boxes}     & 15.2K & 641 & 16.2K \\
\textit{Magazine\_Subscriptions}     & 60.1K & 3.4K & 71.5K \\
\bottomrule
\end{tabular}
\end{table}

\subsection{Prompt Template}
\label{sec:prompting}
We build a promting template with a user prefix and per-item prefixes as below:
\begin{center}
\fbox{%
\begin{minipage}{\textwidth}
\footnotesize
$\vcenter{\hbox{\rule{0.315\textwidth}{0.5pt}}}$
\textrm{[USER PURCHASE HISTORY SEQUENCE]}
$\vcenter{\hbox{\rule{0.315\textwidth}{0.5pt}}}$
\texttt{\textcolor[RGB]{79, 89, 153}{\newline This is the summary of a user’s purchase history.} The first item bought is as follows. \textcolor[RGB]{189, 113, 38}{Title: Funko Marvel Collector Corps Subscription Box, I Am Groot Disney+ Theme, L, Multicolor. Category: SUBSCRIPTION BOXES. Brand: Funko. Price: Unknown. Image: <image>}
\newline The next item bought is as follows. \textcolor[RGB]{189, 113, 38}{Title: Funko Marvel Collector Corps Box. Category: SUBSCRIPTION BOXES. Brand: Unknown. Price: Unknown. Image: <image>} \newline The next item bought is as follows.\newline}
$\vcenter{\hbox{\rule{0.37\textwidth}{0.5pt}}}$
\textrm{[TARGET ITEM SEQUENCE]}
$\vcenter{\hbox{\rule{0.37\textwidth}{0.5pt}}}$
\texttt{\textcolor[RGB]{74, 102, 53}{\newline Title: WWE T-Shirt Club Subscription - Men - 2XL. Category: SUBSCRIPTION BOXES. Brand: Unknown. Price: Unknown. Image: <image>}}
\end{minipage}}
\end{center}

The \emph{same} per-item prefix is used to cue generation. During collation, we pack the full sequence (history + target) and also build a target-only batch for the second tower. We locate the separator token in text space to split hidden states into \emph{history} and \emph{target} segments and apply masking to all history tokens including the image tokens.

\subsection{Inference and Retrieval}
We generate $N_{\text{return}} = 32$ text-only candidates with temperature parameter set to $0.5$, repetition penalty to $1.2$ and using sample decoding, along with their sequence scores. We perform quasi-round-robin BM25 matching over the item corpus using the \textsc{BM25S} \cite{bm25s} implementation; scores are linearly scaled and modulated by $\exp(\epsilon \cdot \text{logprob})$ with $\epsilon = 1/5000$, and then max-pooled across the $N_{\text{pred}} = 20$ generated strings to yield final item rankings.

In our experiments, we retrieve top relevant items based on \textsc{BM25S} (BM25-Sparse) \cite{bm25s}, which is an optimized implementation of the classical BM25 ranking strategy used in information retrieval. Similar to the classic BM25, it scores candidate items by balancing exact term matches against document length normalization, making it effective for matching queries against large corpora. However, \textsc{BM25S} introduces several algorithmic optimizations that make it more suitable for large-scale recommender settings which allow \textsc{BM25S} to deliver slightly more stable and faster retrieval performance, particularly when working with millions of items or when matching against many queries in quasi-round-robin fashion. We report Recall@1, Recall@10, MRR, and NDCG@10 of the generated items following standard practices from prior work.
\begin{figure}[ht!]
    \centering
    \includegraphics[width=0.7\linewidth]{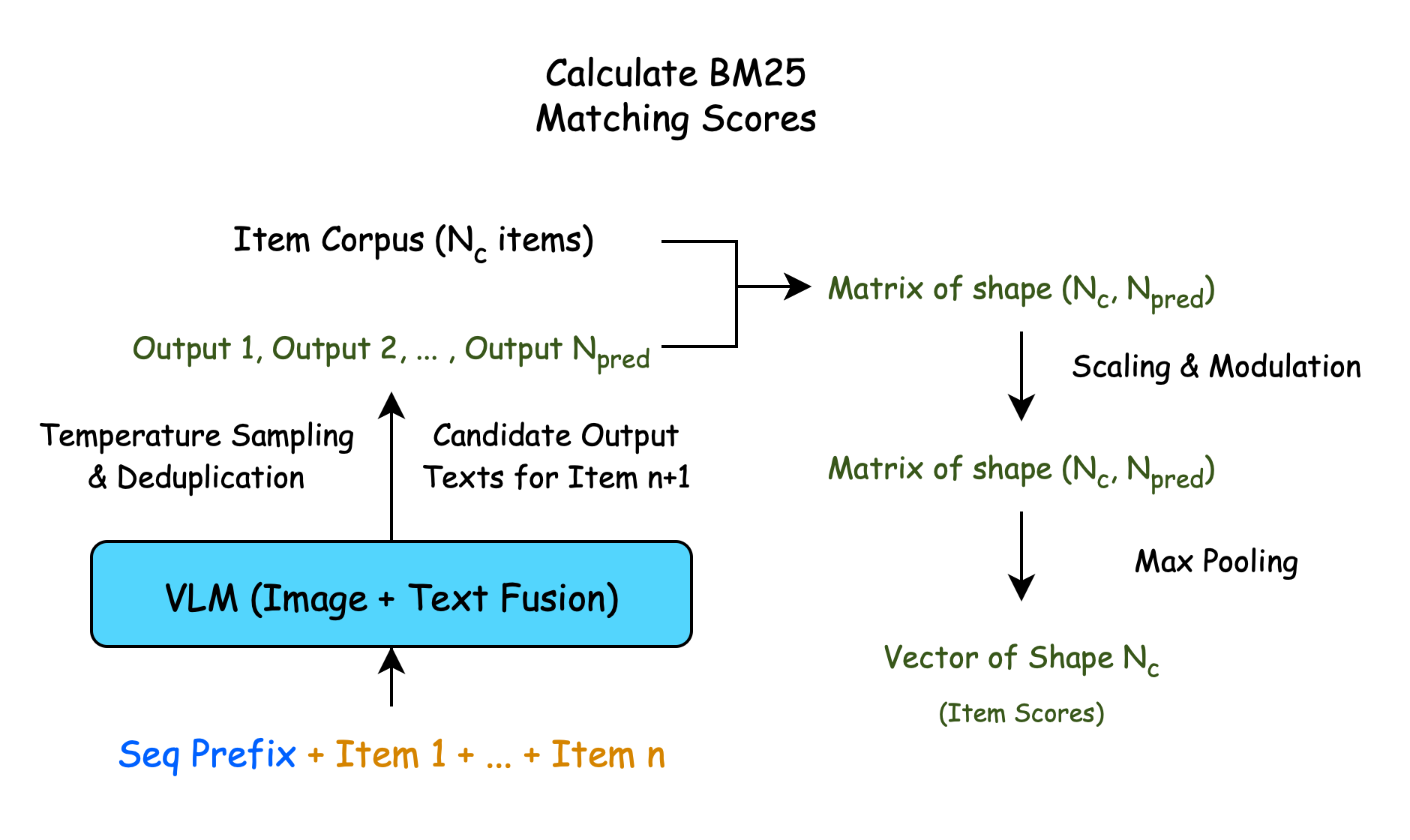}
            \caption{Quasi Round-Robin Best Matching}
            \label{fig:round-robin}
\end{figure}

\subsection{Results and Observations}
Table~\ref{table:vlm_perf} reports the recommendation results on \textit{Subscription\_Boxes} category. Both the VLMs (\textsc{SmolVLM} and \textsc{PaliGemma2}) outperform the text-only model \textsc{SmolLM2} in terms of Recall@1. $\mathrm{NDCG@10}$ increased by $19\%$, $\mathrm{Recall@10}$ increased by ${\sim}40\%$, and $\mathrm{Recall@1}$ was higher by ${\sim}3\times$ in \textsc{PaliGemma2} with respect to \textsc{SmolLM2}. The VLM (\textsc{PaliGemma2}) with more number of parameters consistently outperforms the compact backbone (\textsc{SmolVLM}). Mean Reciprocal Rank remains almost unchanged, suggesting that most of the improvement is concentrated within the top–10 predictions rather than producing significant lift in the position of the first relevant item in the ordering.

The results show remarkable potential of visual context in helping the model to surface the correct product variant (style/color/packaging) into the candidate set even when textual features like title or description are nearly identical as evident for the gain in $\mathrm{Recall@10}$. This is useful for marketplaces where item titles are noisy. In practical e-commerce systems, adding a learned cross-encoder or a re-ranker that fuses product image features with catalog metadata (price, brand, availability) should inherently provide better item rankings thus helping in high rate of conversion (add-to-cart clicks). Since moving from a lightweight to a stronger VLM yields ${\sim}2{\times}$--$3{\times}$ improvements on retrieval-quality metrics, employing \emph{selective routing} seems to be the best approach where the idea is to use a compact model for high quality text+image signals and escalate to a stronger VLM when textual signals are weak. Overall, visual signals substantially improve candidate quality in image-rich recommendation domains like e-commerce providing a clear research path for the future.

\begin{table}[h!]
\begin{center}
\caption{Performance results of fine-tuned Text-based LLMs and VLM models for Sequential Recommendation on \textit{Subscription\_Boxes} category of Amazon Reviews dataset}
\label{table:vlm_perf}
\resizebox{0.8\linewidth}{!}{%
\begin{tabular}{ccccc}
\toprule 
\rowcolor{Gray}
\multirow{1}{*}{} &\multicolumn{1}{c}{} &\multicolumn{3}{c}{\textsc{Model}} \\
\rowcolor{Gray}
\multirow{1}{*}{Category} &\multicolumn{1}{c}{  Metric } &\multicolumn{1}{c}{\textsc{SmolLM2}} &\multicolumn{1}{c}{\textsc{SmolVLM}} &\multicolumn{1}{c}{\textsc{PaliGemma2}} \\
\rowcolor{Gray}
\multirow{1}{*}{}  &\multicolumn{1}{c}{ } &\multicolumn{1}{c}{(Text-Only)} &\multicolumn{1}{c}{(Text+Image)} &\multicolumn{1}{c}{(Text+Image)}\\

\cmidrule(lr){3-5}
\multirow{4}{*}{\textit{Subscription\_Boxes}}
&\multicolumn{1}{c}{NDCG@10} &\multicolumn{1}{c}{0.0086} &\multicolumn{1}{c}{0.0062} &\multicolumn{1}{c}{\textbf{0.0103}}  \\
&\multicolumn{1}{c}{Recall@1} &\multicolumn{1}{c}{0.0013} &\multicolumn{1}{c}{\underline{0.0026}} &\multicolumn{1}{c}{\textbf{0.0045}} \\
&\multicolumn{1}{c}{Recall@10} &\multicolumn{1}{c}{0.0222} &\multicolumn{1}{c}{0.0104} &\multicolumn{1}{c}{\textbf{0.0313}} \\
&\multicolumn{1}{c}{MRR} &\multicolumn{1}{c}{0.0936} &\multicolumn{1}{c}{\underline{0.112}} &\multicolumn{1}{c}{\textbf{0.112}} \\
\bottomrule
\end{tabular}
}
\end{center}
\end{table}

\section{Conclusion and Future Plans}
Our results indicate that visual context substantially improves candidate quality but that rank ordering require additional engineering. We outline a road-map for future improvements that targets quality and reliability of sequential recommendations using visual features with the aim to achieve distinct candidate top-rank gains in multi-modal recommendations and ensure the systems remains robust with varying multi-modal catalogs and evolving user behavior.
\paragraph{\textbf{Model Improvements and Ablations}} Our plan is to systematically study early vs. late fusion strategies of text+image modalities, adapter placement and QLoRA configuration for efficient fine-tuning, temperature and repetition-penalty schedules at inference, and the effects of loss weights $(\alpha,\beta)$ on next-item generation quality. We intend to incorporate category-balanced sampling, calibrate scores and train with list-wise objectives to enhance MRR/Recall@1.
\paragraph{\textbf{Generalization and Scale-Out}} Another prospective direction we intend to explore is to extend our methods to other Amazon Review categories as well as user-behavior domains like tracking click-through rate, add-to-cart, conversions etc. (e.g. Pinterest User Interaction) and benchmark our results across the different datasets to test domain transfer. Through this approach, we aim to provide better generalization across domain-specific recommendation tasks making \textit{PixRec} much more effective and useful in practice.
\paragraph{\textbf{Low-latency Inference}} One way in which we plan improve on the retrieval speed is by replacing lexical matching with embedding-space retrieval using the contrastively-aligned target representations. Moreover, our intention is to cache item embeddings offline and implement \emph{selective routing} where we start with a compact VLM and escalate to a stronger backbone only when textual ambiguity or visual similarity is high. This would not only enhance the inference process but also ensure consistently better multi-modal recommendations in production-grade e-commerce software platforms.

\bibliographystyle{unsrt}  
\bibliography{references}  %%% Remove comment to use the external .bib file (using bibtex).

\end{document}